\documentstyle[12pt]{article}

\begin{document}
\title{Phenomenology of $SU(3) \otimes SU(2) \otimes U(1)$ supersymmetric 
model with Dirac neutrino masses}
\author{N.V.Krasnikov \thanks{E-mail address: KRASNIKO@MS2.INR.AC.RU}
\\Institute for Nuclear Research\\
60-th October Anniversary Prospect 7a,\\ Moscow 117312, Russia}
\date{March,1998}
\maketitle
\begin{abstract}
We consider minimal supersymmetric extension of standard electroweak model 
with Dirac neutrino masses. In such model for significant region of the 
parameters right-handed tau sneutrino is the lightest superparticle and 
right-handed charged tau slepton is the next to lightest superparticle. 
Due to the smallness of the neutrino masses the right-handed tau slepton is 
long-lived particle that changes the standard signatures used in the 
search for supersymmetry at supercolliders. The most striking signatures of 
such scenario is the existence of highly ionizing tracks and excess of 
multilepton events that is similar to the phenomenology of gauge-mediated 
supersymmetry breaking models. 
           
\end{abstract}
\newpage

Supersymmetric electroweak models offer the simplest solution of the gauge 
hierarchy problem \cite{1}-\cite{4}. In real life supersymmetry has to be 
broken and the masses of superparticles have to be lighter than $O(1)$ TeV 
provided the supersymmetry solves the gauge hierarchy problem \cite{4}. 
Supergravity gives a natural explanation for the supersymmetry breaking, 
namely, taking the supergravity breaking into account in the hidden 
sector leads to soft supersymmetry breaking in the observable sector \cite{4}.
For the supersymmetric extension of the Weinberg-Salam model, soft 
supersymmetry breaking terms usually consist of the gaugino mass terms, 
squark and slepton mass terms with the same mass at the Planck scale and 
trilinear soft scalar terms  proportional to the superpotential. Another 
standard assumption is that the ``Minimal Supersymmetric  Standard Model'' 
(MSSM) conserves R-parity. As a consequence of R-parity conservation 
supersymmetric particles can only be produced in pairs and supersymmetric 
particles can't decay into ordinary particles, so the lightest superparticle 
(LSP) is stable. The typical SUSY signature for 
supercolliders involves missing $E_T$ transverse energy  as a signal for SUSY 
particles production \cite{5}. 

In this paper we consider supersymmetric extension of the Weinberg-Salam 
model with nonzero Dirac neutrino masses (the MSSM model with 
Dirac neutrino masses) . We show that in  such model for the significant 
range of the parameters the right-handed sneutrino is the LSP and the charged 
right-handed tau slepton is the next to lightest superparticle. Due to the 
smallness of the  neutrino masses the right-handed tau slepton is 
long-lived particle and it decays outside of the detector that drastically 
changes the standard signatures used in the search for supersymmetry at 
supercolliders. Namely, the most striking signatures for the SUSY search 
in considered model include highly ionizing tracks from long-lived 
right-handed tau sleptons and excess of multi-muon signals.\footnote{
The phenomenology of the considered  model is similar to the 
phenomenology of some gauge-mediated supersymmetry breaking 
models \cite{6} and a model \cite{7} with superweak R-parity violation.}

Consider supersymmetric  $SU(3) \otimes SU(2) \otimes U(1)$  
model with Dirac neutrino masses. The superpotential of the model 
has the form  
\begin{equation}
W = h_{u_i} Q_iH_1\bar{u}_{i} + 
h_{d_i}Q_iH_2\bar{d}_{i} + h_{\nu_{i}}L_iH_1\bar{\nu}_{i} + 
h_{e_i}L_iH_2 \bar{e}_{i} + \mu H_1H_2 .
\end{equation}
Here $Q_i =(u_i,d_i)_L$, $L_i=(\nu_{i}, e_i)_L$, $H_1 =(H_{11}, H_{12})$, 
$H_2 = (H_{21}, H_{22})$, $\bar{u}_i = u^c_{R,i}$, $\bar{d}_{i} = d^c_{R,i}$, 
$\bar{\nu}_{i} = \nu^{c}_{R,i}$, $\bar{e}_i = e^{c}_{R,i}$, 
$H_1H_2 = \epsilon^{ij}H_{1i}H_{2j}$.
The considered model is a minimal generalization of the MSSM, 
the single difference is that neutrinos are massive and Dirac particles. 
The standard assumption of the MSSM is that at GUT scale 
$M_{GUT} \approx 2 \cdot 10^{16} GeV$ soft supersymmetry 
breaking parameters are  universal. For the 
gaugino masses, an account of the evolution from the GUT 
scale to the observable electroweak scale leads to the formula
\cite{8}
\begin{equation}
M_i = \frac{\bar{\alpha}_{i}(M_Z)}{\alpha_{GUT}}m_{\frac{1}{2}} .
\end{equation} 
The gaugino associated with the $U(1)$ gauge group is the lightest 
among the gauginos, and numerically its mass is given by the formula
\begin{equation}
M_1 \approx 0.43m_{\frac{1}{2}} .
\end{equation}
Here $m_{\frac{1}{2}}$ is common gaugino mass at GUT scale.
In the MSSM the right-handed sleptons usually 
are the next to  the lightest superparticles and in 
the neglection of the Yukawa interactions their masses 
are determined by the formula \cite{8}
\begin{equation}
{m}^2_{\tilde{E}_R} \approx m^2_0 + 0.14m^2_{\frac{1}{2}} -0.22\cos{2\beta}
M^2_Z .
\end{equation} 
Here $m_0$ is common squark and slepton mass at GUT scale and 
$\tan(\beta)=\frac{<H_1>}{<H_2>}$. In neglection of the Yukawa interactions 
the right-handed sneutrino masses coincide with $m_0$. As it follows from   
the formulae (2,3) for $m_{\frac{1}{2}} \geq 2.3m_0$ the right-handed 
sneutrino are the lightest superparticles \footnote{As it has been shown 
in ref.\cite{10} righthanded sneutrino with a mass $\sim 2$ GeV is a natural 
candidate for dark matter} 
Especially interesting 
is the particular case when $m_{0} \leq 0.17m_{\frac{1}{2}}$. In this 
case the charged right-handed sleptons are the next to lightest superparticles.
The case $m_0 \ll m_{\frac{1}{2}}$ is theoretically very attractive 
since it allows to solve SUSY flavour changing problem. For the scenario 
when the lightest superparticle is the right-handed sneutrino and the 
right-handed sleptons are the next to lightest superparticles (to be 
precise an account of  nonzero Yukawa interactions makes tau 
right-handed slepton the lightest among right-handed sleptons) 
the right-handed tau slepton is long-lived particle due to the smallness 
of the neutrino masses. 
After the integration over the superfields $H_1$, $H_2$ one can find the 
effective superpotential describing the decay of right-handed sleptons 
into right-handed sneutrino
\begin{equation}
W_{eff} = h_{\nu_{i}}h_{e_j}L^k_iL^l_j \epsilon_{kl}\bar{\nu}_{i}
\bar{e}_{j}\frac{1}{\mu}
\end{equation}

For $m_{\nu_{\mu}} \gg m_{\nu_{e}}$ and $m_{\tilde{\tau}_R} \gg 
m_{\tilde{\nu}_i}$  the 
tau slepton decay width into 
$\tilde{\tau}_R \rightarrow \tilde{\nu}_{\mu,R} \nu_{\tau} \mu, 
\tilde{\nu}_{\mu,R} \nu_{\mu}\tau$
is determined by the formula
\begin{equation}
\Gamma(\tilde{\tau}_{R}) \approx
\frac{1}{192\pi^{3}}\frac{m^3_{\tilde{\tau}_R}}{\mu^{2}}
\frac {m^2_{\tau}m^2_{\nu_{\mu}}}{v^2(\sin(2\beta))^{2}},
\end{equation}
where $v =174 GeV$ and $m_{\tilde{\tau}_R}$ is the 
tau right-handed slepton mass. 
 As it follows from the formula  (5) due to the 
smallness of the muon  neutrino mass the right-handed tau slepton is 
long-lived particle. For instance, for $m_{\tilde{\tau}_R} =
100 GeV$, $\mu=500$Gev, $\sin(2\beta)=0.1$ and $m_{\nu_{\mu}}= 100 eV$
the stau lifetime is $\tau(\tilde{\tau}_R) \sim 0.3$ sek. For
$m_{\nu{\mu}} = 10 eV$ the stau lifetime is $\tau(\tilde{\tau}_R) 
\sim 30$ sec. Such charged particle is long-lived (it decays 
outside of the detector) that changes comletely the signatures for 
the search for supersymmetry at supercolliders. Remember that 
standard signatures used for the search for supersymmetric particles 
at supercolliders in the assumption that LSP is neutral and escapes 
from being registrated at the detector   are events with 
hadronic jets + missing $E_{T}$ + $(n \geq 0)$ isolated leptons. 
Missing $E_{T}$ arises from nonobservation of electrically neutral LSP. 
In our case we shall not have missing transverse energy $E_{T}$ but 
instead as a result of the right-handed sleptons decays we shall 
have two opposite sign long-lived charged sleptons in the SUSY 
event. If such sleptons are nonrelativistic we can distinguish them 
from muons by highly ionized tracks. For relativistic sleptons it is 
difficult to distinguish them from muons so we shall have the excess of 
multi-muon events in the final states with 5 or more isolated muons 
and very little hadronic activity. The standard background with 5 
or more isolated muons is extremely small and the predicted signature is very
clean. It should be noted that in gauge-mediated supersymmetry 
breaking models charged sleptons could be the next to lightest superparticles 
and decay outside of the detector for large region of parameter 
space \cite{6}. Similar situation takes place in a model with superweak 
explicit R-parity violation \cite{7}. The phenomenology these models is very 
similar to the phenomenology of the considered model and it has been 
discussed in refs. \cite{6,9}.       
 
In conclusion let us formulate our main results. We considered the simplest 
generalization of the MSSM model - the MSSM with nonzero Dirac neutrino 
masses. We have shown that for large regions of parameter space 
which are theoretically very attractive the lightest superparticle 
is the  right-handed sneutrino and the next to lightest superparticle 
is right-handed tau slepton. Due to the smallness of t
he tau neutrino 
mass the right-handed tau slepton is long-lived particle and it decays 
outside of the detector that changes comletely the MSSM signatures 
used for the search for supersymmetry at supercolliders.          

I am indebted to V.A.Rubakov for discussion and critical comments. 
The research described in this publication was made possible in part 
by Award No RPI-187 of the U.S. Civilian Research and Development 
Foundation for the Independent States of the Former Soviet Union(CRDF).

\end{document}